\documentstyle[12pt,aps,epsf]{revtex}

\begin{document}

\title{{\Large {\bf THE CHAOTIC PROPERTIES\ OF\ }}$Q-${\Large {\bf STATE\ POTTS
MODEL\ ON\ THE\ BETHE\ LATTICE:\ }}$Q<2${\Large {\bf \ }}}
\author{N.S. Ananikian$^{a,b}$\thanks{
e-mail: ananik@jerewan1.yerphi.am}, S.K. Dallakian$^{a,b}$and B. Hu$^{a,c}$ 
\\
$^{a}${\normalsize Department of Physics and the Centre for Nonlinear
Studies, Hong}\\
{\normalsize Kong Baptist University, Hong Kong, China}\\
$^{b}${\normalsize Department of Theoretical Physics, Yerevan Physics
Institute,}\\
{\normalsize Alikhanian Br.2, 375036 Yerevan, Armenia}\\
$^{c}${\normalsize Department of Physics, University of Houston, Houston, TX
77204,}\\
{\normalsize USA}}
\maketitle

\begin{abstract}
The $Q-$ state Potts model on the Bethe lattice is investigated for $Q<2$.
The magnetization of this model exhibits a complicated behavior including
both the period doubling bifurcation and chaos. The Lyapunov exponents of
the Potts-Bethe map are considered as order parameters. We find a scaling
behavior in the distribution of Lyapunov exponents in fully developed
chaotic case. Using the thermodynamic formalism of dynamical systems have we
investigated the nonanalytic behavior in the distribution of Lyapunov
exponents and located the point of phase transition of the ''chaotic free
energy''.

\end{abstract}

\newpage

\begin{center}
1. INTRODUCTION
\end{center}

The $Q$-state Potts model is one of generalizations of the Ising model
constructed for investigation of the phase transition\cite{M1}. This model
was initially defined for an integer $Q$ but it has also many applications 
for noninteger $Q$. The exact solution of the two-dimensional Potts model
for general $Q$ has been obtained only at the self-dual point by mapping it
into two-dimensional inhomogeneous six vertex model\cite{M2}.

Another exact solution of the Potts model for general $Q$ and coordination
number can be obtained on the Bethe lattice. It is interesting to note that
similar results can be obtained in the $N\rightarrow 1$
limit of $N\times N$ Hermitian matrix model on random graphs\cite{M3}.

The properties of the ferromagnetic and antiferromagnetic Potts models in
the magnetic field have been rigorously considered on the Bethe lattice by
means of a recursion relation\cite{M4}.

The $Q$-state Potts model on the Bethe lattice has been recently
investigated for $Q<2$\cite{M5}. Many physical processes can be formulated
in terms of $Q$-state Potts model when $Q<2$, e.g. the resistor network, dilute
spin glass, percolation and Self Organizing Critical systems\cite{M1,NN3,G1}
It has been shown that when $Q<2$ the $Q$-state Potts model exhibits a
large verity of phase transitions leading to specially modulated and chaotic
phases. These phases are similar to those obtained in Axial Next
Nearest Ising, chiral Potts and three-site interacting Ising models. It is
necessary to point out that contrary to the above mentioned models these phase
in the Potts model ($Q<2$) are obtained without the frustration \cite{G3,G2,A5}.

In this paper we investigate the $Q$-state Potts model on the Bethe lattice
for $Q<2$ using the thermodynamic formalism of multifractals.. The reason of
phase transitions taking place in this model is that the attractor of the
map (Potts-Bethe) which is used for calculation of average quantities has a
complicated dependence on the parameters of the Potts model. Being one
dimensional the Potts-Bethe map exhibits period doubling cascade, chaos,
etc. By using thermodynamic formalism of multifractals one can calculate the
distribution of local Lyapunov exponents of Potts-Bethe map in fully
developed chaotic case. The Lyapunov exponent is not only a good order
parameter for transition to chaos, but also completely characterizes the
system in chaotic states. The general aim of this paper is to find a scaling
in the distribution of local Lyapunov exponents of the Potts-Bethe map.
The multifractal approach allow one to map the problem of computation of
Lyapunov exponent onto thermodynamics of one-dimensional spin model and
interpret scaling in the distribution of Lyapunov exponent as a phase
transition in an one-dimensional spin model\cite{A1,A2,A3,A4}. We obtain the
phase transition temperature by means of numerical calculations of the free
energy of this model.

This paper is organized as follows. The Potts model on the Bethe lattice and
its recursion relation is presented in Sec. 2. In Sec. 3 we discuss the
phase structure of the Potts model on the Bethe lattice. In Sec. 4 we
investigate the Potts-Bethe map for the case of fully developed chaos. By using
the thermodynamic formalism of multifractals, the distribution of local
Lyapunov exponent is obtained and the phase transition is analyzed in terms
of the one-dimensional spin model. Finally, in Sec. 5 we summarize our results
and comment on their implications for the study of other systems.

\begin{center}
2. THE\ POTTS\ MODEL\ ON\ THE\ BETHE\ LATTICE\ AND\ THE\ RECURSION\ 
RELATION\ \end{center}

The Potts model in the magnetic field is defined by the Hamiltonian 
\begin{equation}
{\cal H}=-J\sum_{<i,J>}\delta (\sigma _{i},\sigma _{j})-H\sum_{i}\delta
(\sigma _{i},1),  \label{R1}
\end{equation}

where ${\sigma }_i$ takes the values $1$,$2$,$...$, $Q$, the first sum goes
over all edges and the second one over all sites on the lattice.
Additionally, we use the notation $K=J/kT$, $\ h=H/kT$.

The partition function and single site magnetization is given by

\[
{\cal Z}=\sum_{\{\sigma \}}e^{-{\cal H}/kT}, 
\]

\begin{equation}
M=<\delta (\sigma _{0},1)>={\cal Z}^{-1}\sum_{\{\sigma \}}\delta (\sigma
_{0},1)e^{-{\cal H}/kT},  \label{R2}
\end{equation}
where the summation goes over all configurations of the system.

When the Bethe lattice is cut apart at the central point, it is separated into 
$\gamma $ identical branches. The partition function can be written as
follows 
\begin{equation}
{\cal Z}_{n}=\sum_{\{\sigma _{0}\}}exp\left\{ h\delta (\sigma
_{0},1)\right\} {[g_{n}({\sigma _{0}})]}^{\gamma },  \label{R3}
\end{equation}
where ${\sigma }_{0}$ is the central spin and ${g_{n}({\sigma _{0}})}$ is
the contribution of each lattice branch. The latter is expressed through ${%
g_{n-1}({\sigma _{1}})}$, i.e. the contribution of the same branch
containing $n-1$ generations and starting from the site belonging to the
first generation, 
\begin{equation}
g_{n}({\sigma }_{0})=\sum_{\{\sigma _{1}\}}exp\left\{ K\delta (\sigma
_{0},\sigma _{1})-h\delta (\sigma _{1},1)\right\} {[g_{n-1}({\sigma }_{1})]}%
^{\gamma -1}.  \label{R5}
\end{equation}
Introducing the notation 
\begin{equation}
x_{n}=\frac{g_{n}(\sigma \neq 1)}{g_{n}(\sigma =1)},  \label{R6}
\end{equation}
one can obtain the Potts-Bethe map 
\begin{equation}
x_{n}=f(x_{n-1},K,h),\qquad f(x,K,h)=\frac{e^{h}+(e^{K}+Q-2)x^{\gamma -1}}{%
e^{K+h}+(Q-1)x^{\gamma -1}}.  \label{R7}
\end{equation}
The magnetization of the central site for the Bethe lattice with $n$
generation can be written as 
\begin{equation}
M_{n}=<\delta (\sigma _{0},1)>=\frac{e^{h}}{e^{h}+(Q-1)x_{n}^{\gamma }}.
\label{R8}
\end{equation}

\begin{center}
3.THE\ PHASE\ STRUCTURE\ OF\ THE\ POTTS\ MODEL
\end{center}

Let us consider the magnetization of the central site. In order to achieve
the thermodynamic limit we tend the number of generations to infinity ($n\to
\infty $). The recursion relation (\ref{R7}) converge to a fixed point at
every values of parameters $h,K$ in ferromagnetic case ($K>0$) and has only
one period doubling in the antiferromagnetic case ($K<0$) corresponding to
a rise of antiferromagnetic order in different sublattices for $Q\geq 2$%
\cite{M4}. The situation change drastically for $Q<2$ in contrast to
above case. For systems with $Q$ values in the range $Q<2$ and
with antiferromagnetic interactions or for systems with $Q$ values in the
range $Q<1$ and with ferromagnetic interactions one obtains for $M$ versus $%
h $ bifurcation diagrams with the full range of period doubling cascade,
chaos, etc.\cite{M5}. The Figure 1 shows plots of $M$ versus $h$ for
anti-ferromagnetic case($K=-0.5$ ,$Q=0.8$, $\gamma =3$).

The Potts model has many specially modulated and chaotic phases when $Q<2$ .
the presence of phase transitions is in obvious contradiction to the universality
hypothesis. The transition to chaos is provided by Feigenbaum exponents
which is well known to be an one-dimensional map (Fig. 1). It is interesting to
note that the similar transition has been found in Ising model with
three-site interaction\cite{A5,A6}. The Potts model($Q<2$) and three-site
interacting Ising model have the same universal Feigenbaum exponents.

As was already mentioned the Lyapunov exponents are not only good order
parameters for the transition to chaos but also completely characterize the
system in chaotic states. In the next section we compute the distribution of
Lyapunov exponents by using thermodynamic formalism of multifractals. The
recursion relation of the critical phenomena of the Potts model and
dynamical systems are similar. The thermodynamical formalism of
multifractals connects the thermodynamical quantities of an one-dimensional
spin model and dynamical properties of strange attractors\cite{S,A1,A2,A3,A4}%
.

\begin{center}
4.POTTS-BETHE\ MAP\ IN\ THE\ CASE\ OF\ FULLY\ DEVELOPED\ CHAOS
\end{center}

In this section we impose two restrictions on the parameters in order to
apply the thermodynamic formalism of multifractals to the Potts-Bethe map.
The first one is regards only an odd coordination number $\gamma $ for
getting even function of $x$ of the Potts-Bethe map. The second one is the
requirement that 
\begin{equation}
f(0)=-f(f(0)),  \label{R12}
\end{equation}

from which we obtain the following restrictions on $h$ and $K$ 
\begin{equation}  \label{R31}
\exp (h)=\frac{1-\exp (2K)+2\exp (K)-\exp (K)Q-Q}{2\exp (\gamma K)}.
\end{equation}

The second one is needed for exhibiting the fully developed chaotic behavior
of the Potts-Bethe map in the ($h$,$K$) plane.

For a crisis map (Eqs.(\ref{R7}), (\ref{R31})) we want to describe the
scaling properties of an attracting set for the sequences $x_{n}$ which is
in this case the interval $I$: $[-\exp (-h),\exp (-h)]$ (Fig.2). For an
index $n$, $I$ is partitioned into $2^{n}$ intervals or $n$-cylinders, these
being the segments with identical symbolic-dynamics sequences of length $n$
taken with respect to the maximum point (we follow here Ref. \cite{A2}). The
inverse function of Eq.(\ref{R7}), $h=f^{-1}$, has two branches, $h_{-1}$
and $h_{1}$ as shown in (Fig.2) and the $n$-cylinders are all the nth-order
preimages of $I$. The length of the cylinders is denoted by $l_{\epsilon
_{1},\epsilon _{2},\dots \epsilon _{n}}\equiv h_{\epsilon _{1}}\circ
h_{\epsilon _{2}}\circ \dots \circ h_{\epsilon _{n}}(I)$ where $\epsilon \in
\{-1,1\}$.

Let us consider an one-dimensional Ising-like model. The energy of the given
configuration $\epsilon _1,\epsilon _2,\dots \epsilon _n$ is equal to $\mid
ln{l_{\epsilon _1,\epsilon _2,\dots \epsilon _n}\mid }$. The partition
function $Z(\beta )$ is defined \cite{A1,A2,A3,A4} as 
\begin{equation}  \label{R35}
Z_n(\beta )=\sum_{\epsilon _1,\epsilon _2,\dots \epsilon _n}l_{\epsilon
_1,\epsilon _2,\dots \epsilon _n}^\beta =\sum_{\epsilon _1,\epsilon _2,\dots
\epsilon _n}e^{-\beta \mid ln{l_{\epsilon _1,\epsilon _2,\dots \epsilon
_n}\mid }},
\end{equation}
where $\beta \in (-\infty ,\infty )$ is a free parameter - the inverse
''temperature''. In the limit $n\to \infty $ the sum behaves as 
\begin{equation}  \label{R36}
Z(\beta )=e^{-n\beta F(\beta )},
\end{equation}
which defines the free energy, $F(\beta )$. The partition function $Z(\beta
) $ can be alternatively written

\begin{equation}  \label{R32}
Z(\beta )=\int d\lambda e^{nS(\lambda )-n\lambda \beta }
\end{equation}

The entropy $S(\lambda )$ is the Legandre transform 
\begin{equation}  \label{R37}
S(\lambda )=-\beta F(\beta )+{\lambda }\beta ,
\end{equation}
where the relation between $\lambda $ and $\beta $ is obtained from 
\begin{equation}  \label{R38}
\lambda =\frac d{d\beta }{(\beta F(\beta ))},\quad \beta (\lambda
)=S^{\prime }(\lambda ),
\end{equation}
that have the following meaning: In the limit $n\to \infty $, $%
e^{nS(\lambda )}$ is the number of cylinders with length $l=e^{-n\lambda }$
or, equivalently, with local Lyapunov exponent $\lambda $. The Hausdorf
dimension of the set of points in $I$ having local Lyapunov exponent $%
\lambda $ is $S(\lambda )/\lambda $ .

We point out that the above defined one-dimensional Ising-like model has no
direct physical meaning and is used here for the computation of the
multifractal spectrum $S(\lambda )$ of the local Lyapunov exponents $\lambda 
$ of the map (Eqs.(\ref{R7}), (\ref{R31})).

By using Eqs.(\ref{R7}), (\ref{R31}), (\ref{R35}), (\ref{R36}) we
numerically calculate the free energy at the point $K=-0.5,Q=0.8,\gamma =3$ 
(Fig.3).
One can see from Fig.3 that the free energy has a nonanalytic behavior
around $\beta _{c}\approx -1$, which shows the existence of the first order
phase transition in this regions of $\beta $.

Large deviations of fluctuations of local Lyapunov exponents can be
described by means of $S(\lambda )$. To consider the above results in terms
of the entropy function $S(\lambda )$, let us first discuss the general
view of the entropy function. First of all, it should be positive on
some interval $[\lambda _{min},\lambda _{max}]$. The value $\lambda =ln2$
must belong to that interval, which follows from the fact that the sum of
the lengths of all cylinders on a given level is $1$. Secondly it is often
found that the values of $\lambda _{min}$ and $\lambda _{max}$ are given by
the logarithms of the slopes at the origin .

The precise form of the entropy function is not easy to obtain with great
accuracy. The existence of the first order phase transition implies that 
there
should be a straight line segment in $S(\lambda )$ and the slope of the line
equal to $\beta _c$. This scenario is seen in Fig. 4. The curve in the
figure corresponds to $n=13$. Of course, with the finite-size data, it is
impossible to determine the straight line segment in $S(\lambda )$ and the
straight line will increase with increasing $n$.

\begin{center}
5. CONCLUSION
\end{center}

In this paper we have investigated the $Q$-state Potts model on the Bethe
lattice in an external magnetic field. A strong connection with results from
the theory of dynamical systems including chaos has been pointed out for $%
Q<2 $ .

The local Lyapunov exponents are introduced as order parameters for
characterizing the large variety of phase transitions which take place in
Potts model. For certain values of parameters the distribution of local
Lyapunov exponents are obtained by using the thermodynamic formalism of
multifractal. The scaling in the distribution of the local Lyapunov
exponents are interpreted as a phase transition in the thermodynamics of the
one dimensional Ising-like model. This phase transition is analyzed in terms
of the ''temperature'' and local Lyapunov exponents $\lambda $.

We remark that similar behavior have been found in three-site interacting
Ising model in the Husimi three\cite{A6,S1}.

The non-integer ($Q<1$) valued Potts model is connected to the gelatation
and vulcanization of branched polymers\cite{V1}. A dense Mandelbrot set 
of Fisher's zeroes for the non-integer valued Potts model can be obtained 
as the three-site antiferromagnetic interaction Ising model\cite{E}.
 Few monolayers of polymers
are described by higher-dimensional maps\cite{V3}.The thin films of branched
polymers can be regarded as the critical behavior of period p-tuplings in
the coupled 1D maps\cite{E0}. The investigation of modulated phases and
chaotic properties in polymers would be discussed in further publications.

This work is supported in part by grants from the Hong Kong Research Grants
Council (RGC),the Hong Kong Baptist University Faculty Research Grant (FRG)
and (N.A.,S.D.) by grant INTAS-96-690.

\newpage

\begin{figure}[b]
\epsfysize=15cm\centerline{\epsfbox{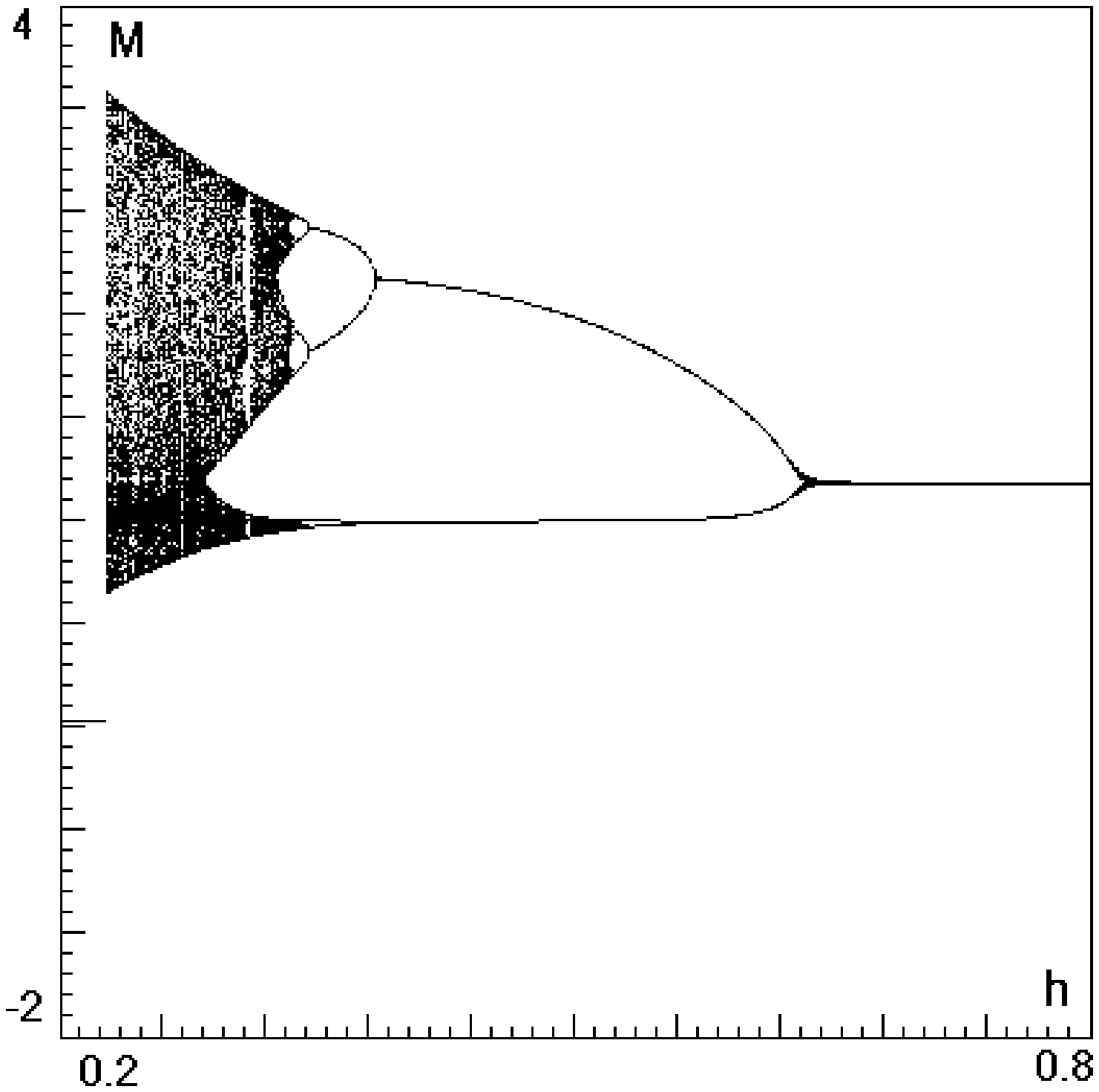}} \vspace{6pt}
\caption{Plot of $m$ - magnetization versus $h$ - external magnetic field 
($K=-0.5$ ,$Q=0.8$, $\gamma $ = 3).}
\label{figone}
\end{figure}

\newpage

\begin{figure}[b]
\epsfysize=15cm\centerline{\epsfbox{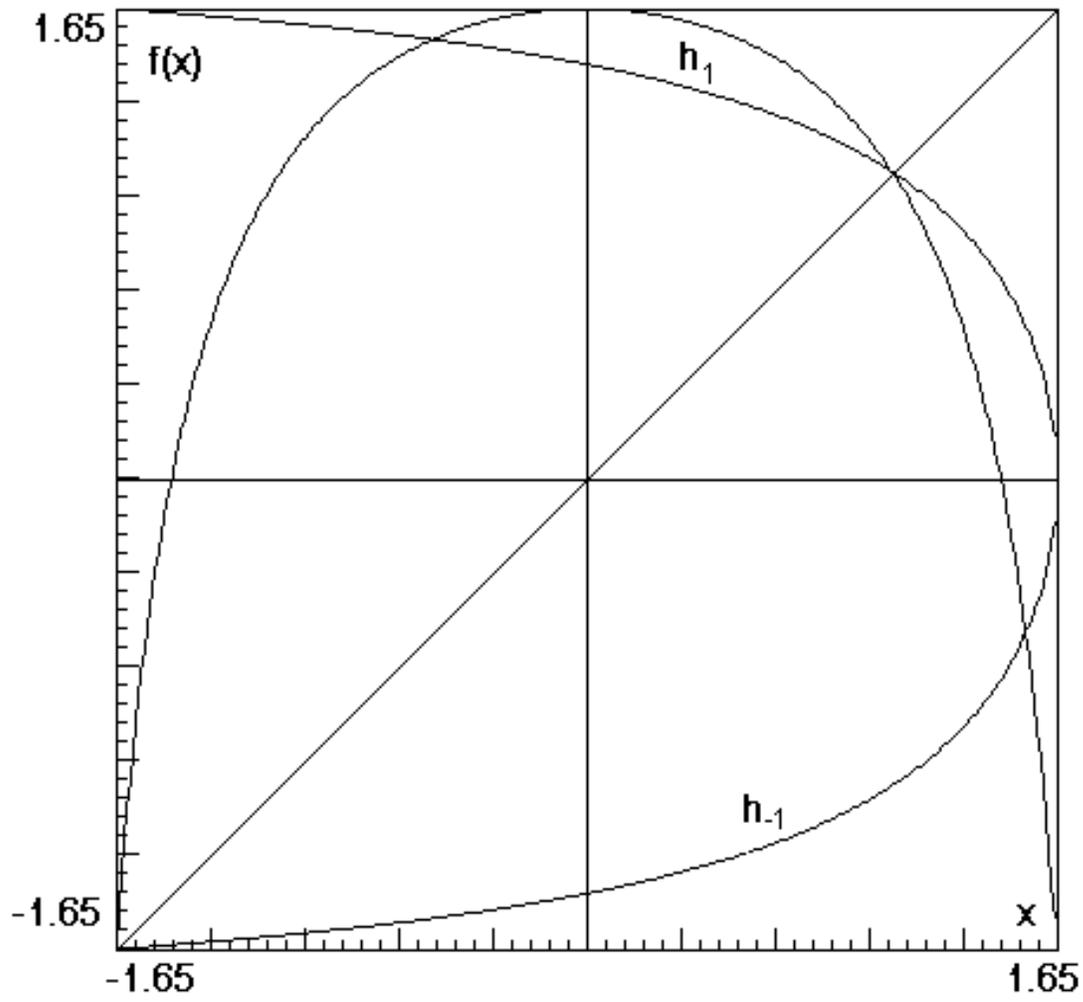}} \vspace{6pt}
\caption{The function of Eq.(6) for $K=-0.5$, $Q=0.8$, $h=0.23$ 
($\gamma $ = 3).}
\label{figtwo}
\end{figure}

\newpage

\begin{figure}[b]
\epsfysize=15cm\centerline{\epsfbox{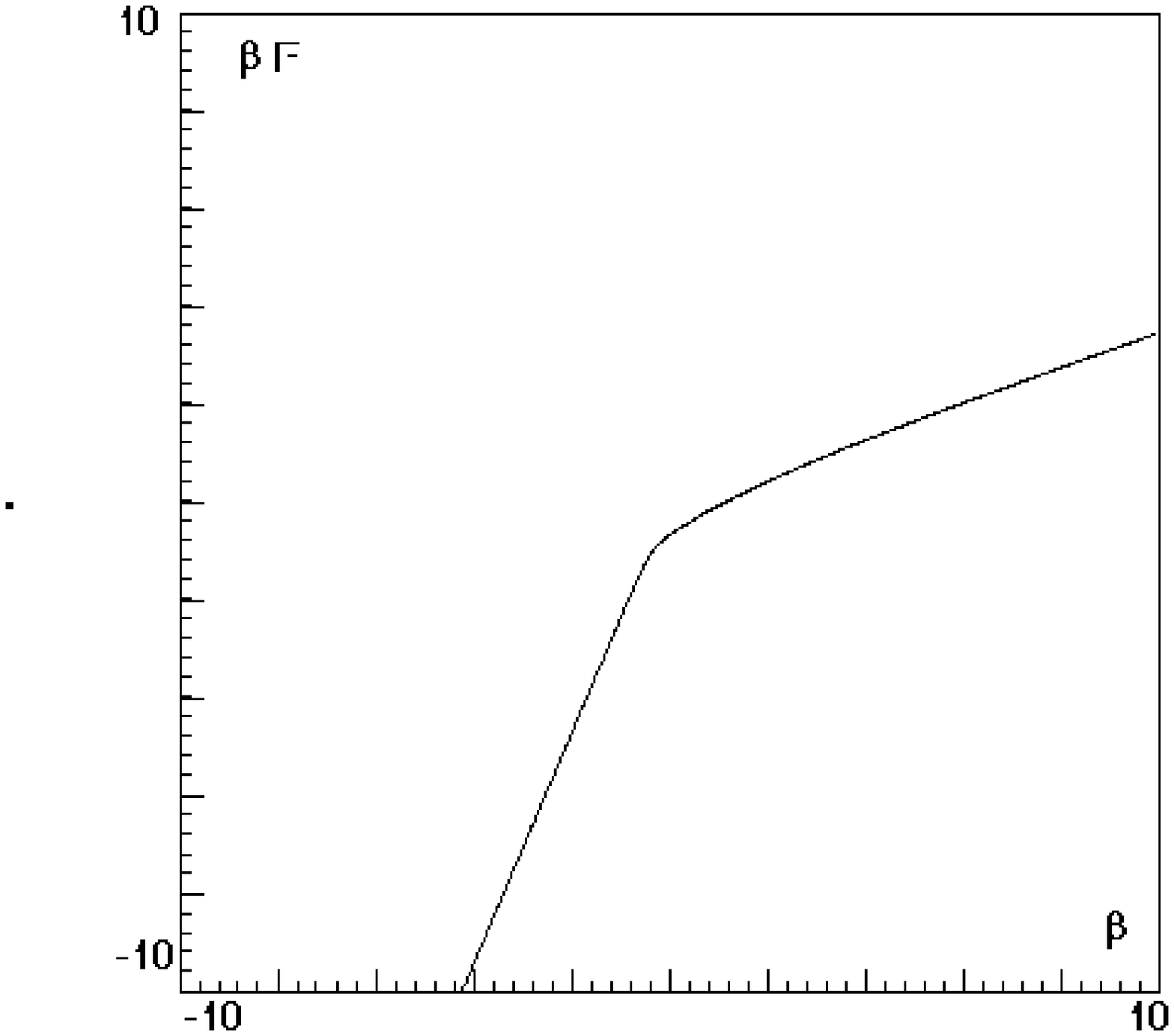}} \vspace{6pt}
\caption{$F(\beta )$, for $K=-0.5$, $Q=0.8$, $h=0.23$ ($\gamma $ = 3).}
\label{figthree}
\end{figure}

\newpage

\begin{figure}[b]
\epsfysize=15cm\centerline{\epsfbox{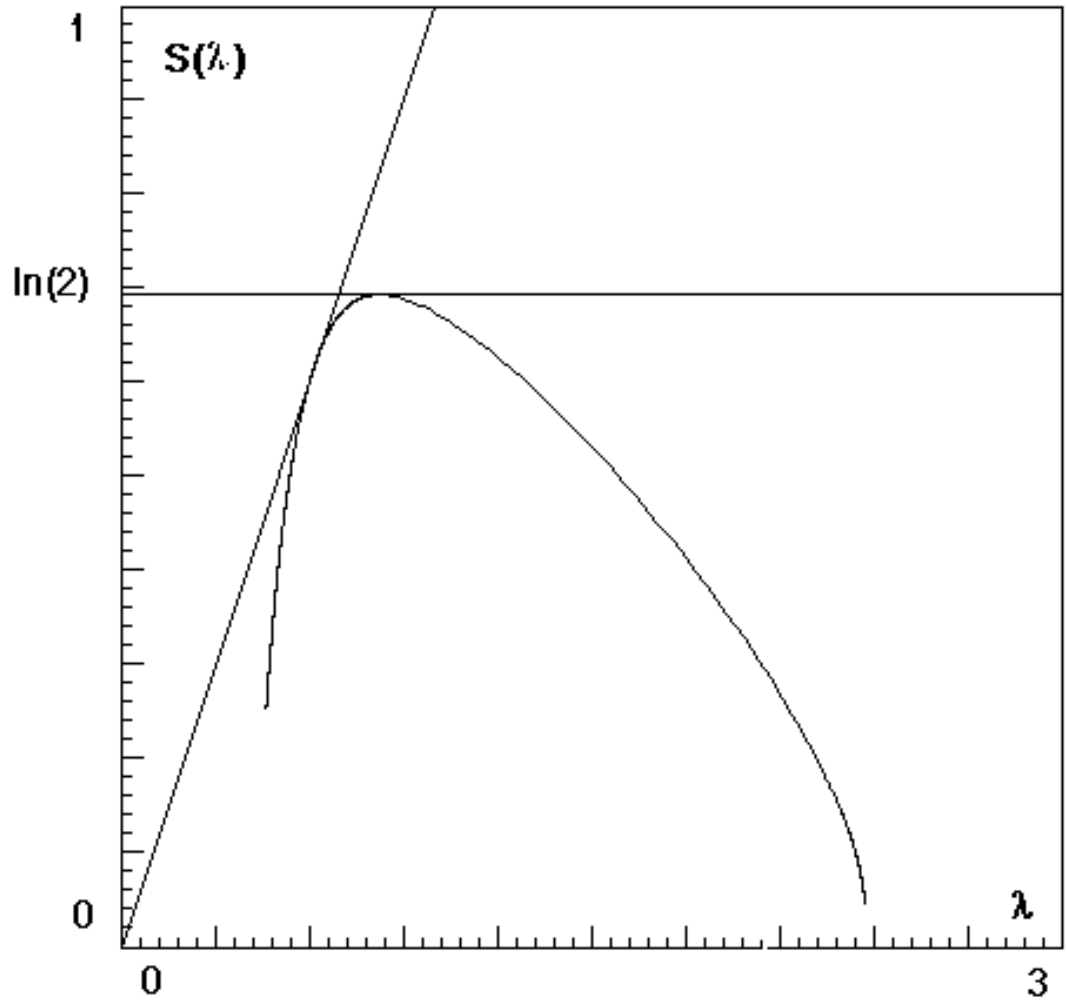}} \vspace{6pt}
\caption{$S(\lambda )$ corresponding to $n$ = 13 for $K=-0.5$, $Q=0.8$, $%
h=0.23$ ($\gamma $ = 3).}
\label{figfour}
\end{figure}

\end{document}